\documentclass[12pt]{iopart}

\usepackage{graphicx}

\begin{document}

\title{Open problems in theory of nuclear open quantum systems}

\author{N. Michel$^1$, W. Nazarewicz$^{2-4}$, J. Oko{\l}owicz$^5$ and M. P{\l}oszajczak$^6$}

\address{$^1$ Department of Physics, Post Office Box 35 (YFL), FI-40014 University of Jyv{\"a}skyl{\"a}, Finland\\
$^2$ Department of Physics and Astronomy, University of Tennessee, Knoxville, Tennessee 37996, USA\\
$^3$ Physics Division, Oak Ridge National Laboratory, Oak Ridge, Tennessee 37831, USA\\
$^4$ Institute of Theoretical Physics, University of Warsaw, ul. Ho\.za 69,
PL-00-681 Warsaw, Poland\\
$^5$ Institute of Nuclear Physics, Polish Academy of Sciences, Radzikowskiego 152, PL-31342 Krak\'ow, Poland\\
$^6$ Grand Acc\'el\'erateur National d'Ions Lourds (GANIL), CEA/DSM - CNRS/IN2P3,
BP 55027, F-14076 Caen Cedex, France}
\ead{witek@utk.edu, ploszajczak@ganil.fr}

\begin{abstract}
Is there a connection between the branch point singularity at the
particle emission threshold and the appearance of cluster states which
reveal the structure of a corresponding reaction channel? Which nuclear
states are most impacted by the coupling to the scattering continuum?
What should be the most important steps in developing  the theory that
will truly unify nuclear structure and nuclear reactions? The common
denominator of these questions is the continuum shell-model approach to
bound and unbound nuclear states, nuclear decays, and reactions.
\end{abstract}

\maketitle

\section{Introduction}
The nuclear shell model was proposed almost sixty years ago \cite{Gop49,Hax49}. Soon afterwards, the interacting shell model (ISM) was developed by Lane \cite{Lan55}, Kurath \cite{Kur56}, and others (see \cite{Cau05} for a recent review). The ISM-based description  of an evolution of the nucleonic coupling scheme from the $LS$ to $jj$ coupling with increasing mass number \cite{Lan55,Kur56,Wil57}, provided foundations of modern nuclear structure theory and helped to understand and categorize  a wealth of data on nuclear levels,  moments, collective excitations, electromagnetic and  $\beta$ decays, and various particle decays \cite{Boh75}. 

In its traditional form, ISM describes the nucleus as a closed quantum system: nucleons occupy bound, hence well localized,  single-particle orbits of an infinite (e.g., harmonic oscillator) potential and are isolated from the environment of unbound scattering states that are not square integrable. Since the scattering continuum is not considered explicitly, the presence of branch points (decay thresholds) and double-poles of the scattering matrix ($S$-matrix) is  neglected. The divide between the discrete states and the scattering continuum, i.e., the focus on one or another,  has unfortunately become a kind of paradigm. In the long term, this has led to an artificial separation of nuclear structure from nuclear reactions, and hindered a deeper understanding of nuclear properties. Indeed, many structural properties of the nucleus are determined by means of nuclear collisions. Hence, the knowledge of nuclear structure depends on nuclear reactions and {\em vice versa}, and this cries out for a unified theoretical framework. 

The first attempt in this direction came from Feshbach  \cite{Fes58,Fes62} who expressed the collision matrix of the optical model using matrix elements of the Hamiltonian. This development gave a strong push to the ISM approach to nuclear reactions \cite{Bre59,Rod61,Mac64,Fan61} (see introduction in Ref. \cite{Bar73} for a detailed historical account) and led to various formulations of the continuum shell model (CSM) \cite{Mah69,Bar77,Phi77,Rot78,Ben99,Vol06}. A modern version of CSM in the Hilbert space, the Shell Model Embedded in the Continuum (SMEC) \cite{Ben99,Rot06}, provides a unified description of the structure and reactions with up to two nucleons in the scattering continuum using  realistic ISM Hamiltonians.  Nevertheless, the fully symmetric treatment of bound and scattering states in the multiparticle wave function is still too ambitious a goal \cite{Oko03}.

A different attempt to formulate the ISM for open quantum systems has been proposed recently \cite{Mic02,Idb02,Mic03,Mic04} within the Berggren ensemble \cite{Ber68}. The resulting complex-energy open quantum system extension of the ISM, the Gamow Shell Model (GSM),  can be conveniently formulated in the Rigged Hilbert Space (Gel'fand triple) \cite{Gel61,Mau68}, which encompasses Gamow states \cite{Gam28,Gur29}, and is suitable for extending the quantum mechanics into the domain of time-asymmetric processes (e.g. decays). The GSM offers a fully symmetric treatment of bound, resonance, and scattering single-particle states but, until now, has been primarily used  in the context of nuclear structure.  (For a recent review of the complex-energy shell model, see Ref. \cite{Mic09}.)

In this paper, we shall draw attention to salient threshold effects related to generic properties of branch points and $S$-matrix poles that are expected to impact structural properties of nuclei. In Sec.~\ref{avoided}, we shall use the phase rigidity indicator to demonstrate effects of the branch point at the particle emission threshold and the exceptional point (the double-pole of the $S$-matrix) on the configuration mixing in the continuum. Section~\ref{instability} discusses the stability of ISM eigenfunctions in the neighborhood of the reaction threshold. Finally, in Sec.~\ref{unification} we outline necessary developments  to achieve a unified theory of structure and reactions offering a symmetric treatment of bound, resonance, and scattering states.

\section{Salient near-threshold phenomena}
At low excitation energies, well-bound nuclei can be considered as closed quantum systems, well described by the standard ISM or its modern versions such as the no-core shell model \cite{Nav96,Nav00,Rot07}. Moving towards drip lines, or higher in excitation energy, the continuum coupling  becomes gradually more important, changing the nature of weakly bound states. (Properties of unbound states are impacted by couplings to reaction channels.) In this regime, the  chemical potential has a  magnitude similar to the pairing gap; hence, the system is dominated by many-nucleon correlations which no longer  can be considered as small perturbations atop the average potential  \cite{Dob07}. Many-body states in neighboring nuclear systems with different proton and neutron numbers  become interconnected via continuum, forming correlated domains (clusters) of quantum states. At present, little is known about the basic features of such clusters, e.g. a typical `cluster size' and its dependence on binding properties of participating states. 
\begin{figure}[htb]
\begin{center}
\includegraphics[width=0.8\linewidth]{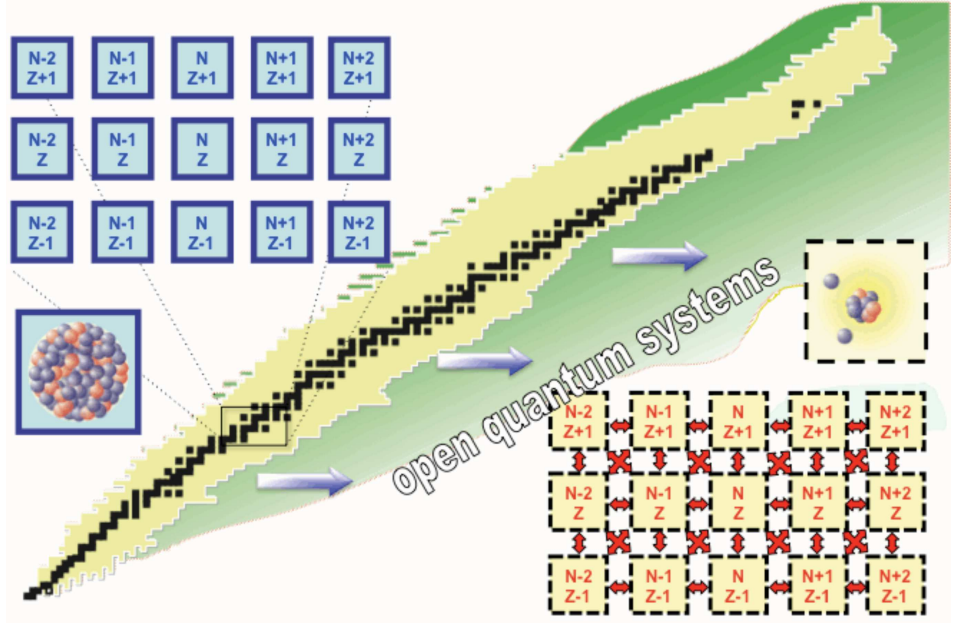}
\caption{At low excitation energy, well-bound nuclei close to the stability valley can be described as  closed quantum systems. When moving towards the particle drip lines, nuclei with different proton and neutron numbers  become interconnected via continuum space, i.e., reaction channels, forming correlated domains of quantum states.} 
\label{network}
\end{center}
\end{figure}

In standard network problems, like the percolation problem \cite{Sta85,Bot02}, different phases of correlated domains  are formed depending on the mean network activity with respect to some threshold. Analogously, one expects that the key  to understanding the formation of correlated domains of quantum states is a  near-threshold behavior of a single network member. In the correlation-dominated regime of a nuclear system, the percolation threshold(s) should be identified with  reaction threshold(s). 

What can be said about properties of many-body systems  in the narrow range of energies around the reaction threshold? Are those properties independent of any particular realization of the Hamiltonian? Below, we shall try to address these  questions in the framework of SMEC, but essentials of this discussion are valid in the GSM approach as well.

\subsection{The main ingredients of the continuum shell model}

To facilitate further discussion, we shall now recall fundamentals of SMEC. 
The Fock space of an $A-$particle system consists of the set of square-integrable functions 
${\cal Q}\equiv \{\Phi_i^{(A)}\}$, internal space described by the standard ISM,  and the space of embedding scattering states ${\cal P}\equiv \{\zeta_E^{c}\}$. These two orthogonal and complementary subspaces of the Fock space are obtained by the projection formalism. In one-particle continuum problems, reaction channels `$c$' correspond to the motion of an unbound nucleon in a state $(lj)$ relative to the  daughter nucleus in some ISM eigenstate $\Phi_k^{(A-1)}$. An open quantum system description of `internal' dynamics, i.e., in ${\cal Q}$, includes couplings to the 
`environment' of scattering states and decay channels and is given by the energy-dependent effective Hamiltonian:
\begin{eqnarray}\label{HQQ} 
{\cal H}_{QQ}(E)=H_{QQ}+W(E), 
\end{eqnarray}
where $H_{QQ}$ is the ISM Hamiltonian (the closed quantum system Hamiltonian) in the internal space ${\cal Q}$ and 
\begin{eqnarray}\label{coupl}
W(E)=H_{QP}G_P^{(+)}(E)H_{PQ}
\end{eqnarray} 
is the energy-dependent term describing both virtual particle excitations and irreversible decays to the environment of reaction channels. In the latter expression,  $G_P^{(+)}(E)$ is the Green's function for the motion of a single nucleon  in the continuum, and $E$ is the nucleon's energy  (i.e., the scattering energy). The external mixing of two ISM states $i$ and $j$ due to $W(E)$ consists of the Hermitian principal value integral $W^R(E)$ describing virtual excitations to the continuum and the anti-Hermitian residuum $W^I(E)=-(i/2){\bf VV}^T$ in a dyadic form that represents decay out of the internal space ${\cal Q}$. The $M\times\Lambda$ matrix ${\bf V}=\{V^c_i\}$ denotes the amplitudes connecting the ISM state $\Phi_i$ ($i=1,\cdots,M$) to the reaction channel $c$ ($c=1,\cdots,\Lambda$) \cite{Oko03}. 

At energies below the lowest reaction threshold, the effective Hamiltonian is Hermitian, i.e., $W^I=0$. Above the first  threshold, the non-Hermitian part of ${\cal H}_{QQ}(E)$ describes the irreversible decay from the internal space. The effective Hamiltonian in this case becomes complex symmetric. Each eigenstate of  ${\cal H}_{QQ}(E)$ is coupled to states in neighboring nuclei via a network of reaction channels, either closed or open. The contribution of different reaction channels to the total continuum coupling is highly non-uniform and spans over a considerable range of excitation energies \cite{Oko07}. 
Since the Hamiltonian (\ref{HQQ}) depends explicitly on the energy, it is highly non-linear. Moreover, the continuum-coupling term generates effective many-body interactions in the internal space, even if it is originally  two-body in the full space. 

Effects of the resulting many-body interactions have been  studied by considering  the continuum-coupling correction to the binding energy  for the chains of oxygen and fluorine isotopes \cite{Oko07,Luo02}. It has been found that the induced many-body interactions explain a significant shift in the neutron drip line for  fluorine isotopes as compared to the oxygens. Another explanation, explicitly  invoking effective three-body interaction, has recently been suggested in Ref.~\cite{Ots09}. These two scenarios could be difficult to distinguish.

Eigenstates of the open quantum system Hamiltonian ${\cal H}_{QQ}(E)$ are biorthogonal. The left $|\Psi_j\rangle$ and right $|\Psi_{\tilde j}\rangle$ eigenstates have complex conjugate wave functions. The Hermitian conjugation of  ${\cal H}_{QQ}$ switches the  left and right vectors. The orthonormality condition in the biorthogonal basis can be written as  $\langle\Psi_{\tilde j}|\Psi_k\rangle=\delta_{jk}$. 

The diagonalization of the effective SMEC Hamiltonian can be achieved by means of an orthogonal transformation. The resulting eigenvalues  are real if all reaction channels are closed. Above the  threshold, the transformation becomes non-unitary:
\begin{eqnarray}
\label{transf}
{\Phi}_i \longrightarrow \Psi_j = {\sum}_{i}^{} b_{ji}{\Phi}_i 
\end{eqnarray}
and it yields complex eigenvalues. Physical resonances can be identified with narrow poles of the 
$S$-matrix 
\cite{Sie39,Ber68,Mic09}, or using the Breit-Wigner approach, which leads to a fixed-point condition \cite{Vol06,Oko03,Mad05}. 

\subsection{How to monitor effects of branch points and avoided level crossings in the complex energy plane?}\label{avoided}

The energy-dependent term (\ref{coupl}) describing the coupling of closed quantum system eigenfunctions $\Phi_i$ to the environment of scattering states does not act uniformly on all ISM states. A useful indicator of effects caused by an anti-Hermitian coupling term $W^I(E)$ on biorthogonal eigenfunctions $\{\Psi_k\}$ of the effective Hamiltonian  is the phase rigidity \cite{Lan97,Bro03,Bul06}:
\begin{eqnarray}
r_j=\frac{\langle\Psi_{\tilde j}|\Psi_j\rangle}{\langle\Psi_j|\Psi_j\rangle}~ \ ,
\label{rig}
\end{eqnarray}
given by the ratio of biorthogonal and Hermitian norms of an eigenfunction. The indicator  (\ref{rig}) can also be written in a form \cite{Bul06}:
\begin{eqnarray}
r_j=e^{2i\theta_j}\frac{\int dr \left(|{\rm Re}{\bar \Psi}_j(r)|^2-|{\rm Im}{\bar \Psi}_j(r)|^2\right)}{\int dr \left(|{\rm Re}{\bar \Psi}_j(r)|^2+|{\rm Im}{\bar \Psi}_j(r)|^2\right)} ~ \ ,
\label{rig1}
\end{eqnarray}
which better illustrates its physical meaning. In this expression, the angle $\theta_j$ arises from a transformation of $\Psi_j$ so that ${\rm Re}{\bar \Psi}_j$ and ${\rm Im}{\bar \Psi}_j$ are orthogonal and it characterizes the degree to which the eigenfunction $\Psi_j$ is complex.

Phase rigidity varies between 1 and 0. It equals 1 for bound state eigenfunctions. For unbound states, the condition $r_j$=1 means that the non-Hermitian coupling $W^I(E)$ exerts a negligible effect on the structure of an eigenfunction, i.e., its biorthogonal and Hermitian norms are identical. The energy variation of $r_j$ is an internal property of a considered open quantum system and measures a mutual influence of the neighboring resonances. 

\begin{figure}[htb]
\begin{center}
\includegraphics[scale=0.42]{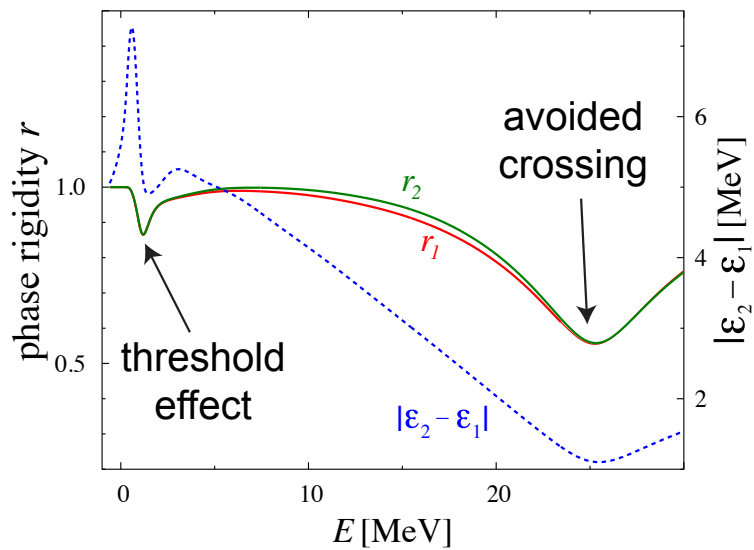}
\caption{The energy dependence of the phase rigidity for the wave functions of the  $0^+_1$ and $0^+_2$ SMEC states of $^{16}$Ne. The dashed curve shows the difference between energies of these states in the complex energy plane.} 
\label{rigid}
\end{center}
\end{figure}
Figure~\ref{rigid} shows the phase rigidity as a function of the scattering energy  $E$ for the two lowest-energy $0^+$ levels of the effective Hamiltonian of $^{16}$Ne. These SMEC calculations are performed in a $p_{1/2}, d_{5/2}, s_{1/2}$ model space. For $H_{QQ}$ we use the ZBM Hamiltonian \cite{Zuk68}.  The residual coupling $H_{QP}$ between the internal space and the surrounding continuum is generated by the contact force: $H_{QP}=H_{PQ}=V_0\delta(r_1-r_2)$ with $V_0=-1100$ MeV fm$^3$.  The $0^+$ ISM eigenstates are coupled through
the common {\em physical} one-proton emission channels $[^{15}{\rm F}(I^{\pi})\otimes {^1{\rm H}}(lj)]_{E'}^{J^{\pi}}$ with $I^{\pi}=1/2^+, 5/2^+$, and $1/2^-$ which have  thresholds at $E=0$ (elastic channel), 0.67 MeV, and 2.26 MeV, respectively. (Each physical channel, specified by a states $I^{\pi}$ and $J^{\pi}$ in daughter $^{A-1}X$ and parent $^AY$ nuclei, respectively, may correspond to a number of different mathematical channels characterized by different $(lj)$-values of a nucleon which are allowed by the coupling of $I^{\pi}$ and $J^{\pi}$.)
These are all possible one-proton  emission channels in $^{16}$Ne, described in ZBM space.

One can notice two local variations of the phase rigidity for $0_1^+$ and $0_2^+$ eigenfunctions. The first one appears close to the elastic channel threshold whereas the second one, at higher energies,  indicates the avoided level crossing between $0_1^+$ and $0_2^+$ eigenvalues in the complex energy plane.  Avoided level crossings are traces of exceptional points \cite{Kat95,Zir83,Hei91} in the parameter space of the effective Hamiltonian  \cite{Her03,Oko09}.  
The dashed curve in Fig. \ref{rigid} shows a distance in the complex energy plane between the two $0^+$ eigenvalues. The minimum in $\Delta\varepsilon=|\varepsilon_1-\varepsilon_2|(E)$ coincides with the minimum in $r(E)$ at a position of the avoided crossing.  However, neither maximum nor minimum of $\Delta\varepsilon$ coincides with the minimum of the phase rigidity around the elastic reaction threshold.

\begin{figure}[htb]
\begin{center}
\includegraphics[scale=0.42]{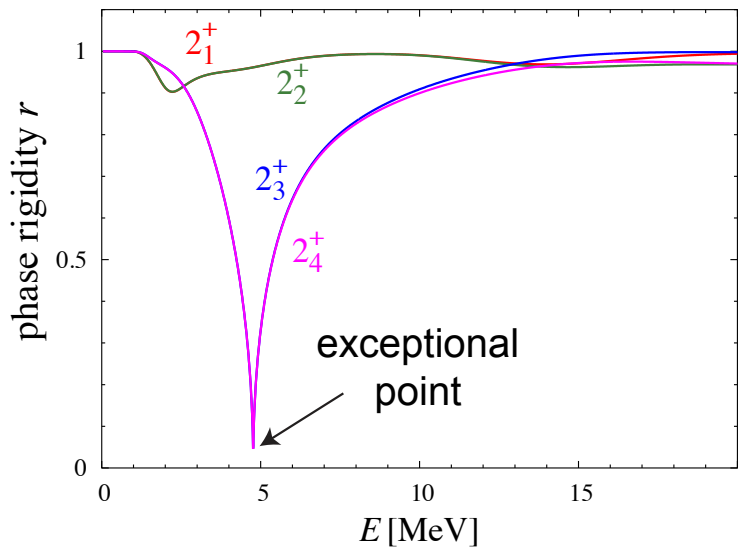}
\caption{The energy dependence of the phase rigidity for lowest $2_i^+$ ($i=1,\cdots,4)$ eigenfunctions  of the  SMEC Hamiltonian in $^{16}$Ne. The $2_3^+$ and $2_4^+$ eigenvalues form an exceptional point at $E \simeq 4.8$ MeV.} 
\label{rigid3}
\end{center}
\end{figure}

A striking example of the configuration mixing caused by the coupling term $W(E)$ is shown in Fig. \ref{rigid3}. The phase rigidity  
for $2_i^+$ ($i=1,\cdots,4$) states in $^{16}$Ne is plotted here for  $V_0=-1357$ MeV fm$^3$. The $2^+$ states are interconnected via the coupling to the inelastic channel: $[^{15}{\rm F}(5/2^{\pi})\otimes {^1{\rm H}}(lj)]_{E'}^{2^{+}}$. One can clearly see that at this value of the continuum coupling strength there appears the degeneracy (exceptional point) of $2_3^+$ and $2_4^+$ levels at which the phase rigidity  drops abruptly to 0. The remaining $2^+$ eigenfunctions are mere spectators of this violent wave function rearrangement.
At low energies, close to the branch point at the channel threshold ($E$=0.67 MeV), one can see a strong mixing of $2_1^+$ and $2_2^+$ eigenfunctions.

\subsection{Near-threshold instability of the wave function}\label{instability}

\begin{figure}[htb]
\begin{center}
\includegraphics[scale=0.6]{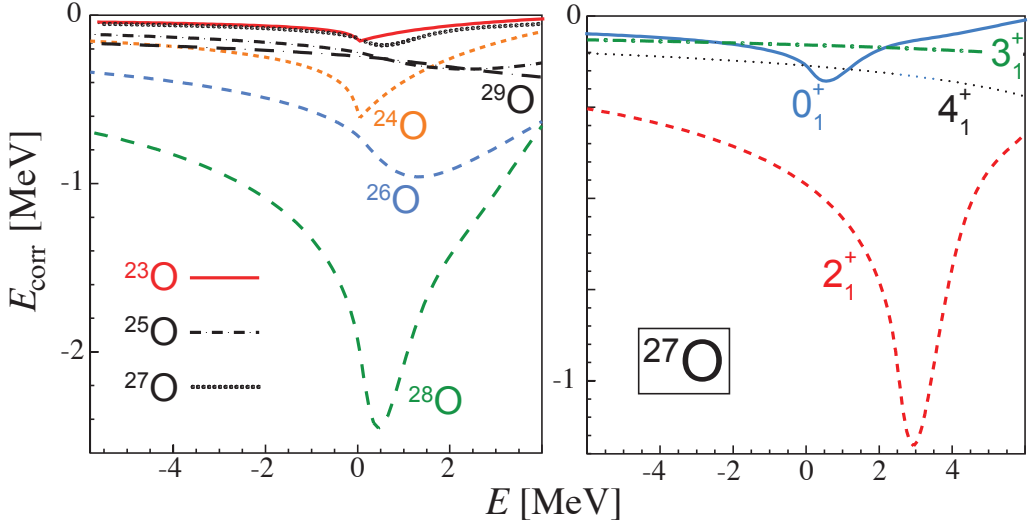}
\caption{The continuum-coupling energy correction $E_{\rm corr}(E)$. The value $E$=0 corresponds to the first (one-neutron emission) threshold. For $H_{QQ}$, we take USD Hamiltonian for the $sd$ shell \cite{Wil84} and KB' interaction for the $pf$ shell \cite{Pov81}. The cross-shell interaction is the G-matrix \cite{Kah69}. The coupling to continuum is given by the Wigner-Bartlett interaction with the strength $V_0=-414$ MeV fm$^3$.
Left: ground state configurations of  $^{23-29}$O. Only a contribution from
a single channel corresponding to a neutron coupled to the ground state in the daughter $^{A-1}$O nucleus is included.
Right: the ground state configuration of $^{27}$O. Contributions from couplings to various positive parity states  in $^{26}$O are indicated.
} 
\label{ccorr}
\end{center}
\end{figure}

Both the branch points associated with reaction thresholds and the avoided level crossings are essential elements of the configuration mixing in open quantum systems. Abrupt variations of the phase rigidity as a function of the energy for certain eigenstates of the effective Hamiltonian are indicative of  wave function  instability. Another indicator of such instabilities could be the continuum-coupling energy correction, $E_{\rm corr}(E)=\langle\Phi_i|W(E)|\Phi_i\rangle$, to the ISM eigenvalue $\langle\Phi_i|H_{QQ}|\Phi_i\rangle$. Figure~\ref{ccorr} (left)  shows the continuum-coupling correction 
for the ground states of oxygen isotopes calculated for a single channel corresponding to a neutron coupled to the ground state in the daughter nucleus $^{A-1}$O. 

One can see that the continuum-coupling energy correction does depend on wave functions involved in the coupling matrix element, i.e., the wave function $\Psi_i$ in the parent nucleus $[N,Z]$ and the channel wave function $[_ZX_{N-1}(I^{\pi})\otimes{^1{\rm n}}(lj)]^{J^{\pi}}_{E'}$. The correction
$E_{\rm corr}$ varies strongly  in a broad interval of energies around the neutron emission threshold: 
from $\pm4$ MeV in $^{28}$O to $\pm2$ MeV in $^{26}$O and $^{24}$O. 
As a result of the centrifugal barrier,  a maximum of $|E_{\rm corr}|$ is shifted above the threshold if the angular momentum $l$ of a neutron is different from zero.  For  $l$=0, the first derivative of $|E_{\rm corr}(E)|$  is discontinuous at the neutron emission threshold. A similar shift can be caused by a Coulomb barrier if the continuum coupling involves  charged particle emission channels. 
Another interesting observation is the odd-even staggering of $E_{\rm corr}$ \cite{Luo02}: a blocking of the virtual scattering to the particle continuum by an odd nucleon diminishes the continuum correction to the binding energy in  odd-$N$ nuclei.

Figure \ref{ccorr} (right) shows individual contributions to the continuum-coupling energy correction in 
$^{27}$O coming from couplings to various reaction channels  $[_8{\rm O}_{18}(I^{\pi})\otimes{^1{\rm n}}(lj)]^{J^{\pi}}_{E'}$.  The largest contribution to $E_{\rm corr}$ comes from the coupling to the lowest physical inelastic channel: $[_8{\rm O}_{18}(2_1^+)\otimes {^1{\rm n}}(lj)]^{3/2^+}$. The same wave appears in the  elastic channel
$[_8{\rm O}_{19}(3/2_1^+)\otimes {^1{\rm n}}(lj)]^{0^+}$ contribution to $E_{\rm corr}$   in the neighboring nucleus $^{28}$O (see Fig.~\ref{ccorr} on the left).

The angular momentum dependence and charge dependence of the continuum-coupling correction around the particle-emission threshold 
can be explained in the same way as 
a characteristic threshold behavior of scattering and reaction cross sections \cite{Wig48} (Wigner cusp).  Due to the unitarity of the $S$-matrix and the resulting flux conservation, Wigner's threshold effect has a broad impact on various channel wave functions due to channel coupling  \cite{Baz57,New59,Mey63,Baz66,Hat78}. Hence, the branch point singularity at the opening of a given  reaction channel can induce non-local correlations between wave functions in distant channels. This effect has been found experimentally, for example, in the coupling of the analogous channels in $(d,p)$ and $(d,n)$ reactions \cite{Moo66}. These channel-channel correlations are yet another manifestation of the eigenfunction correlations induced by the presence of branch points or exceptional points. 

Recently, it has been found that a threshold law, similar to the Wigner's law for scattering and reaction cross-sections, holds for  one-nucleon overlap amplitudes  \cite{Mic07}, which probe  shell occupancies; hence, they characterize  many-body correlations. An immediate consequence of this finding is that the influence of the branch point on the configuration mixing in weakly bound mirror states will be different, even for  identical separation energies in mirror states \cite{Mic09a}.

The size of the energy correction is a measure of  the mixing of unperturbed ISM wave functions due to the coupling to the reaction channel(s), either open or closed. For ISM wave functions below all reaction thresholds, only the Hermitian part of the coupling term $W(E)$ acts. For unbound ISM wave functions, the competition of Hermitian $W^R$ and anti-Hermitian $W^I$ parts is an essential ingredient of the formation mechanism of exceptional points in the spectra of resonances. As shown above, $|E_{\rm corr}|$ is both state- and system-dependent, i.e., the continuum coupling does not induce a global instability of all ISM eigenstates close to the particle emission threshold. On the contrary, for those ISM wave functions which are susceptible to the continuum coupling, the near-threshold instability is  seen predominantly in a single pair of unperturbed ISM states. Other ISM states with the same quantum numbers remain spectators even though all of them are interconnected by the coupling to the same reaction channel(s).  Both avoided level crossings (or exceptional points) and branch points associated with reaction thresholds induce the configuration mixing of continuum states. For weakly bound states, only branch point(s) can contribute to the configuration mixing.

\begin{figure}[htb]
\begin{center}
\includegraphics[scale=0.75]{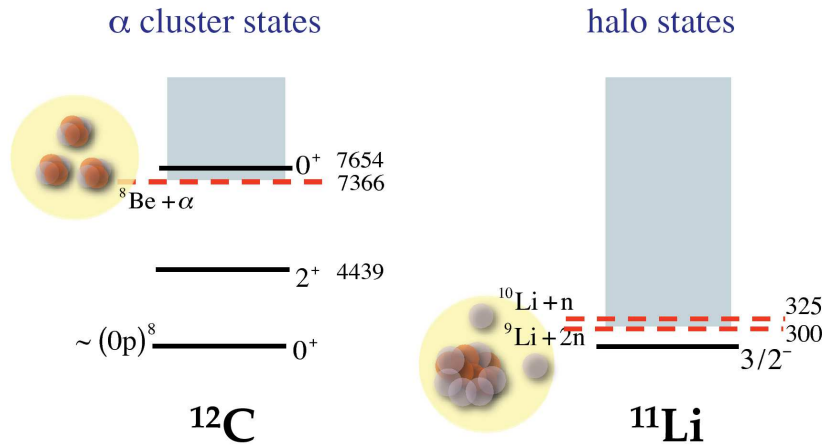}
\caption{Cluster states close to decay thresholds.
Left: Excited 0$^+$ state in $^{12}$C (Hoyle state) just above the 
$^{8}$Be+$\alpha$ threshold. Right: Subthreshold halo ground state of $^{11}$Li.
} 
\label{clusters}
\end{center}
\end{figure}

In general, the mixing mechanism associated with the coupling of ISM eigenstates to the reaction channel(s) enhances the similarity of a bound, near-threshold ISM state $\Psi(J^{\pi})$ with the decay channel $[\Psi^{(d)}(I^{\pi})\otimes({\tau_z}lj)]^{J^{\pi}}$,
i.e., increases the corresponding spectroscopic factor $\langle {\tilde \Psi(J^{\pi}})||a_{lj}^+||\Psi^{(d)}(I^{\pi})\rangle^2$, where
$\Psi^{(d)}$ is the  wave function of the  daughter
system. (The tilde symbol above bra vector signifies that the complex conjugation in the dual space affects the angular part and leaves the radial part unchanged.) 
This `alignment' of a near-threshold eigenstate of the effective Hamiltonian with the corresponding reaction channel has been demonstrated in the approximation of a one-nucleon continuum \cite{Cha06}. The phenomenon is based on three general principles: (i) the branch point nature of the reaction channel threshold, (ii) the unitarity of the $S$-matrix, and (iii) the $2\times2$ matrix structure in the eigenfunction mixing. These principles neither invoke any particular interaction in the internal space nor any special structure of eigenfunctions involved in the mutual coupling. Hence, the alignment of near-threshold eigenfunctions of the effective Hamiltonian with the reaction channel should be present in any open quantum many-body system  which has complex reaction channels. In particular, the entanglement of ISM wave functions due to their coupling to a common particle emission channel should be responsible for the omnipresence of clustering phenomena in states close to their cluster decay thresholds. Indeed, as illustrated in Fig.~\ref{clusters}, $\alpha$-cluster states have been found among the ISM wave functions close to the $\alpha$-particle emission thresholds in, for example, $^{12}$C, whereas correlated two-neutron pair has been seen in the halo ground state of $^{11}$Li close to the two-neutron emission threshold. 

The threshold features are generic: they are neither accidental nor arise from  particular properties of the nuclear Hamiltonian.
This universality forms a foundation of Ikeda systematics \cite{Ike68} which can be extended naturally to halo phenomena, see Fig.~\ref{clusters}. In this formulation,  Borromean systems correspond to an example of  clustering where the cluster itself is not  a self-bound system.

We are still far away from a quantitative understanding of the organization of ISM eigenfunctions near the particle emission threshold. The structure of the ground state wave function of $^{11}$Li is a particularly interesting case-study due to an intricate sequence of low-lying two- and one-neutron emission thresholds which are separated by less than 30 keV (!). This approximate degeneracy of two different emission thresholds may have important consequences on the structure of halo two-neutron pair. The  relative weights of configurations, such as 
$[^9{\rm Li}(3/2^-)]\otimes[{\rm n}^2(0^+)]$, corresponding to the two-neutron continuum of $^9$Li, and
$\{[^9{\rm Li}(1/2^-)]\otimes[{\rm n}(1/2^-)]\}^{1^+}\otimes[{\rm n}(1/2^-)]$,
$\{[^9{\rm Li}(1/2^-)]\otimes[{\rm n}(1/2^+)]\}^{1^-}\otimes[{\rm n}(1/2^+)]$, or
$\{[^9{\rm Li}(3/2^-)]\otimes[{\rm n}(1/2^-)]\}^{2^+}\otimes[{\rm n}(1/2^-)]$, associated with an unbound $^{10}$Li, reflects the influence of these two almost degenerate branch points on the structure of the ground state wave function of $^{11}$Li.

The exceptional points, whose dramatic effects on the phase rigidity can be seen in Fig. \ref{rigid3}, are also generic in Hamiltonian systems. They arise as a result of an interplay between Hermitian and non-Hermitian parts of the continuum coupling term $W(E)$ for energies above the elastic channel threshold. The number of exceptional points depends only on some basic features of the system, such as its dimensionality and quantum integrability \cite{Duk09}. On the contrary, the precise position of any exceptional point in the spectrum of eigenvalues of the energy-dependent effective Hamiltonian depends on the choice of the ISM Hamiltonian $H_{QQ}$, the strength of the coupling term $W(E)$, and the energy $E$ of the system. Hence, its finding provides the test of both the effective interaction and the configuration mixing in any OQS. 

It has been found recently that low-energy exceptional points  appear for realistic values of coupling to the continuum and hence could be accessible experimentally \cite{Oko09}. At low excitation energies, they could be seen, for example, as individual peaks associated with a jump by $2\pi$ of the elastic scattering phase shift. Also in the neighborhood of an exceptional point for avoided crossing of resonances, this characteristic imprint of the scattering phase shift remains,  which gives a real chance that traces of the exceptional could actually be searched for in reaction studies.

Complex and biorthogonal eigenstates of ${\cal H}_{QQ}(E)$ provide a natural basis in which one can express the resonant part of any scattering wave function:
\begin{eqnarray}
\Upsilon^c_E=\zeta^c_E+\sum_ja_j{\tilde \Psi}_j ~ \ ,
\label{ups}
\end{eqnarray}
where $a_j=\frac{\langle\Psi_{\tilde j}|H_{QP}|\zeta^c_E\rangle}{(E-E_j)}$
and
${\tilde \Psi}_j\equiv(1+G_P^{(+)}H_{PQ})\Psi_j$.
Dominant contributions to $\Upsilon^c_E$ inside  an interaction region  are given by eigenfunctions $\Psi_j$ of ${\cal H}_{QQ}$, i.e.,
\begin{eqnarray}
\Upsilon^c_E\sim\sum_ja_j\Psi_j ~ \ .
\end{eqnarray}
Both branch points (reaction thresholds) and double poles of the $S$ matrix (exceptional points) lead to a non-separable entanglement  of two eigenstates of the effective Hamiltonian which manifests itself in a singular behavior of matrix elements of any operator which does not commute with ${\cal H}_{QQ}$ \cite{Oko09,Duk09,Oko09b}. These unusual features are yet another facet of the profound change of the nature of ISM states close to the reaction thresholds and avoided resonance crossings which are responsible for clustering and strong mixing in OQSs.

\subsection{Future developments: happy marriage of nuclear structure and nuclear reactions}\label{unification}

To understand the formation of a network of correlated states, e.g.,  the chain of isotopes from a well-bound $^9$Li to an unbound $^{12}$Li, a structure of the Hoyle resonance, and the radiative capture reactions, such $^{12}$C($\alpha,\gamma$)$^{16}$O, a theoretical framework is required that would provide a unified approach to nuclear structure and reactions. Such a theoretical framework could be developed based on the GSM \cite{Mic02,Idb02,Mic03,Mic04}. This model, formulated in the Rigged Hilbert Space \cite{Gel61,Mau68} and using a complete Berggren ensemble \cite{Ber68}, is a generalization of the nuclear ISM for a description of  bound states, resonances, and many-body scattering continuum. In a nuclear structure application, solutions of the GSM can be found by diagonalizing a complex-symmetric Hamiltonian matrix. The `dimensional catastrophe' in GSM when increasing the number of active particles  is much more serious than in the standard ISM because each single-particle continuum state in the Berggren ensemble becomes a new shell in the many-body GSM formulation. This acute problem has been  alleviated by recent progress in the generalization of the Density Matrix Renormalization Group  \cite{Whi92,Whi93,Car99}
method to non-Hermitian, complex-symmetric matrix problems \cite{Rot06a,Rot09}.

Significant progress has also been made in  applications of realistic interactions in GSM \cite{Hag06}. Finally, powerful techniques for a selection of physical resonances, based on the overlap method, have been developed \cite{Mic02,Mic03}.  Still, much work is needed to develop appropriate effective interactions which would allow  a systematic study of the structure of nuclear states in long isotopic or isotonic chains: from the valley of stability towards  particle drip lines. 

So far, most applications of the GSM have addressed  nuclear structure phenomena. Further progress can only be achieved if the method could be fully extended to  reaction problems. This can be achieved by the coupled-channel formulation of the scattering problem:
\begin{eqnarray}
\int dr'\sum_{c'}\sum_{m,n} \langle c;r|\Psi_n\rangle\left[\langle \Psi_{\tilde n}|H-E\delta_{nm}|\Psi_m\rangle\right]\langle\Psi_{\tilde m}|c';r'\rangle \Omega_{c'}(r')=0 ~ \ .
\label{cceq}
\end{eqnarray}
A similar approach, based on the Resonating Group Method \cite{Wil58,Wil59,Tan78},
has been developed in the no-core shell-model approach to nuclear scattering problems \cite{Qua08,Qua09}. In the above equation,
$H_{nm}\equiv\langle\Psi_n|H|\Psi_m\rangle$ is the GSM Hamiltonian matrix and $O_{c;n}\equiv\langle c;r|\Psi_n\rangle$ are overlaps of the reaction channel and the GSM many-body state. Formally, Eq.~(\ref{cceq}) takes the same form independently of the choice of reaction channels; the whole complication is hidden in the overlaps $O_{c;n}$. Their calculation is a formidable task for the GSM approach. It requires a complicated change of the coordinate system which makes the calculation difficult  for heavier projectiles, even  for reactions with  two-body asymptotic conditions. Any future progress in the application of GSM to nuclear reactions is ultimately related to the progress in the development of new algorithms for  fast calculation of overlaps.

\section{Acknowledgments}
We wish to thank J. Rotureau for useful discussions.
This work was supported in part by the U.S.
Department of Energy under Contract No. DE-FG02-96ER40963 (University
of Tennessee) and by the CICYT-IN2P3
cooperation.

\end{document}